\documentclass[11pt,a4paper]{article}  
\usepackage{amsmath}  
\usepackage[T1]{fontenc}  
  
\usepackage{caption}
\usepackage{caption}
\usepackage{tabularx}
\usepackage{tabularx,ragged2e}

\usepackage{array}
\usepackage{amsfonts}
\usepackage{xcolor}

\usepackage{xfrac} 

\usepackage{cite}
\usepackage[english]{babel}
\usepackage{multicol}

\usepackage{verbatim}
\usepackage[figurename=Fig.]{caption}
\usepackage[a4paper, lmargin=0.08\paperwidth, rmargin=0.08\paperwidth, tmargin=0.08\paperheight, bmargin=0.08\paperheight]{geometry} 

\begin{document} 

\section*{Plasmon-plasmon interaction and the role of buffer in epitaxial graphene micro-flakes}

 M. Shestopalov$^1$, V. D\v edi\v c$^1$, M. Rejhon$^{1,2}$,  B. Morzhuk$^1$, V. C. Paingad$^4$, I. Mohelský$^3$, F. Le Mardelé$^3$, P. Kužel$^4$, M. Orlita$^{3,1}$ and J. Kunc$^1$ 
\\

\noindent 1) Charles University, Faculty of Mathematics and Physics, Ke Karlovu 5, 12116 Prague, Czech Republic

\noindent 2) Department of Chemical and Biomolecular Engineering, Tandon School of Engineering, New York University, 6 MetroTech Center, Brooklyn, NY 11201, USA

\noindent 3) LNCMI-EMFL, CNRS UPR3228, Univ. Grenoble Alpes, Univ. Toulouse, Univ., Toulouse 3, INSA-T,  Grenoble and Toulouse, France

\noindent 4) Institute of Physics of the Czech Academy of Sciences, Na Slovance 2, 18221 Prague 8, Czech Republic

\section*{Abstract}
We investigate the origin of the translational symmetry breaking in epitaxially grown single-layer graphene. Despite the surface morphology of homogeneous graphene films influenced by the presence of mutually parallel SiC surface terraces, the far-infrared magneto-plasmon absorption is almost independent of the angle between the probing light polarization and the orientation of terraces. Based on a detailed analysis of the plasmon absorption lineshape and its behavior in the magnetic field, supported by confocal Raman mapping and atomic force microscopy, we explain this discrepancy by spontaneously formed graphene micro flakes. We further support our conclusions using data collected on artificially created graphene nanoribbons: we recognize similar plasmon origin in artificial ribbons and naturally formed grains. An unexpectedly large plasmon resonance redshift was observed in nanoribbons. In a hydrogen-intercalated sample (which does not contain the buffer), this redshift is quantitatively taken into account by a plasmon-plasmon interaction. In non-intercalated samples featuring a buffer layer, this redshift is due to an interplay between the plasmon-plasmon coupling and Coulomb screening by the buffer-induced interface states. This model determines the density of interface states in good agreement with experimentally reported values.
\\

\noindent Keywords: epitaxial graphene, magnetoplasmon, plasmon, surface morphology, core-shell model, graphene nanoribbons

\vskip 10mm

\section{Introduction}
A spatial confinement and enhancement of the electromagnetic field, chemical and electrostatic tuneability, low losses, and far-to-mid-infrared response are the driving forces of graphene surface plasmon (GSP) research~\cite{Goncalves2016}. The confinement of the electromagnetic field is determined by graphene's unique dispersion relation of collective charge excitations, plasmons~\cite{Wunsch2006,Hwang2007,Brey2007}. Graphene plasmons show a relatively long lifetime~\cite{Koppens2011,Yan2012}. In contrast to conventional plasmonic metals, graphene's natural ability to form purely two-dimensional layers without clustering is another advantage, predetermining it for nanophotonic device miniaturization. Hence, graphene is a promising material for photonic and optoelectronic applications.

Despite its strong light-matter interaction, the momentum mismatch between light dispersion and plasmon dispersion prohibits any finite-energy plasmon excitations in unperturbed graphene. To make such two-dimensional plasmons optically active, the translational symmetry needs to be broken, or the light dispersion needs to be modified by a high index of refraction in the attenuated total reflection (ATR) geometry~\cite{Goncalves2016}. 
\\
The standard technique of translational symmetry breaking is the spatial confinement in graphene, i.e., by periodic graphene structures in the forms of stripes or discs~\cite{Vasic2013, Wu2014, Fei2015, Wang2016, Xia2019,Padmanabhan2020} or by deposition of metallic stripes on graphene sheet~\cite{Jadidi2015}.  
The spectral position of plasmonic absorption in graphene nanoribbons (GNR) can be tuned by GNR widths and the number of graphene layers~\cite{Wu2014}. Uniquely in graphene, for its low density of states, the tunability is also allowed by electrical~\cite{Koppens2011,Ju2011,Vakil2011, Rodrigo2017} and optical~\cite{Gorecki2018} gating. 
\\
Apart from other graphene fabrication methods, the epitaxial graphene on silicon carbide (SiC) provides an intrinsic translational symmetry breaking with no need for ATR geometry or artificially made nanostructures yet maintains its tunability. Such intrinsic plasmons appear in a large area (few mm$^2$) unpatterned epitaxial graphene~\cite{Crassee2012,PaingadAFM2021}. Crassee et al.~\cite{Crassee2012} assign the origin of spatial confinement to morphological structures identified using Atomic Force Microscopy (AFM). The sources of confinement could be SiC step bunching or wrinkles in graphene created due to the thermal relaxation after epitaxial graphene growth and hydrogen intercalation~\cite{Cambaz2008,Emtsev2009,Forti2011}. The intriguing small degree of polarization sensitivity pointed towards the rough shape of the SiC step edges or randomly oriented wrinkles~\cite{Crassee2012}. It was inferred by Paingad~\cite{PaingadAFM2021} that the plasmonic carrier confinement could be related to the dimensions of single-crystalline graphene flakes surrounded by dead layers with defects independently of the anisotropy imprinted by the SiC terraces. However, besides such generic interpretations, the actual reason for forming nearly isotropic plasmons is unknown.
\\
In this work, we resolve the origin of the intrinsic spatial confinement in epitaxial graphene. We study plasmon and magneto-plasmon resonances in samples with a higher degree of structural anisotropy caused by SiC step bunching. The highly anisotropic step bunching and an unchanged poor degree of polarization enable us to unambiguously exclude the SiC step bunching as a cause of the intrinsic confinement. Instead, we show a good correlation between the far-infrared plasmon resonance and the characteristic graphene flake size determined by Raman mapping. We confirm our conclusions by studying artificially made graphene nanoribbons (GNR). The plasmon resonance in GNR shows unexpectedly large redshift compared to plasmon dispersion in graphene, modified core-shell model, and even compared to the exact solution of Maxwell equations. We show that the redshift is caused by the plasmon-plasmon interaction and by charging of the buffer layer in argon-grown graphene. We also demonstrate that the latter redshift contribution vanishes after hydrogen intercalation, eliminating the buffer layer. The proposed model allows for determining the spectral density of states in the buffer layer.

\section{Materials and Methods}

\subsection*{Epi-graphene fabrication and patterning}
Epitaxial graphene was grown in an induction RF furnace by thermal decomposition of SiC. We grew graphene on Si-face SiC(0001) nominally on-axis oriented wafers. The wafer (II-VI Inc., semi-insulating) was diced into 5$\times$5~mm$^2$ samples and heated in a graphite crucible at 1~atm and flow of 30~slph of Ar atmosphere. The quasi-free-standing single-layer graphene (QFSLG), sample 1, was grown in two steps. First, the buffer layer forms at 1550$^\circ$C/5~min. The sample is removed from the furnace, and opened graphite crucible is used for hydrogen intercalation. The hydrogen intercalation takes place as a second growth step. We described details of hydrogen intercalation in our previous work~\cite{KuncAIPAdvances2018}.\\
We also study three GNR samples. The two samples used for the fabrication of GNRs were grown at 1650$^\circ$C for 5~min (samples 2 and 3). These growth conditions lead to the formation of a Single-Layer Graphene (SLG), which consists of the buffer layer and one graphene layer. These two samples were not hydrogen intercalated, as the plasmon resonance is visible even in graphene with a buffer layer in nanofabricated ribbons. This contrasts with the large-scale graphene grown in argon, where we observe no plasmon resonance unless hydrogen intercalation is performed. The fourth sample is a patterned, hydrogen-intercalated buffer in the form of GNRs. The growth procedure is identical to the unpatterned QFSLG, sample 1; however, we also lithographically prepared an array of GNRs, distinguishing sample~4 from sample~1. We removed graphene grown on the C-face by reactive ion etching (RIE) for the purpose of transmission measurements. The RIE (Oxford Instruments, PlasmaPro 80) was done in oxygen at 40~mTorr, 20~sccm, 50~W for 60~s (DC bias 250~V).\\
GNRs were patterned by electron beam lithography (Raith 150Two). The polymethylmethacrylate positive resist AR-P 679.04 (Allresist) was spun directly on graphene at 4000~rpm for 60~s, including spin-off at 6000~rpm for 5~s to avoid resist accumulation at the sample's edges. The resist was baked at 150$^\circ$C for 15~min to improve lateral resolution of electron beam lithography and resist development. The sample was exposed by 0.29~nA electron beam at the dose 300~$\mu$Ccm$^{-2}$ (line spacing and step size were 40~nm). The exposed resist was developed for 3~min in AR-P 600-56 (Allresist); development was stopped in isopropyl alcohol for 30~s and cleaned in deionized water for 30~s. The developed samples were etched by RIE for 40~s (other abovementioned parameters). An array of GNRs covers the samples' area $4\times4\mathrm{mm}^2$. The fabricated GNRs are 370, 2500 and 3270~nm wide for samples 2, 3, and 4, respectively. The samples' parameters are summarized in Tab.~\ref{tab:studiedsamples}.

\begin{table}[h!]
\centering
\caption{Studied samples.}
\begin{tabular}{l c c c c} 
 \hline
 \hline
   &Sample 1& Sample 2 & Sample 3  & Sample 4 \\ [0.5ex] 
 \hline
Substrate &  \multicolumn{4}{c}{semiinsulating-SiC(0001)} \\
Polytype &4H & 6H & 6H &  6H\\
Growth ambient   &\multicolumn{4}{c}{Argon} \\
Grown & buffer & SLG & SLG & buffer\\
H$_2$ intercalation &Yes&No&No& Yes \\
GNR patterning & No & Yes &  Yes& Yes \\
Final structure &QFSLG & SLG-GNR 370 nm & SLG-GNR 2500 nm  & QFSLG-GNR 3270 nm \\
\hline
\end{tabular}
\label{tab:studiedsamples}
\end{table}

\subsection{Characterization techniques}
Low temperature (4.2~K) polarization-resolved transmittance spectra in magnetic fields up to 16~T (top-loaded superconductive coil) were measured in High Magnetic Field Laboratory in CRNS, Grenoble (France) using vacuum Fourier transform infrared (FTIR) spectroscope Vertex 80v (Bruker) equipped with a polyethylene supported holographic polarizer and silicon bolometer. We used a silicon beamsplitter and a mercury lamp as a light source. The spectral resolution was 4~cm$^{-1}$.  Room temperature far-infrared transmittance was also measured in vacuum FTIR spectroscope Vertex 80v with polypropylene (PP) polarizer and silicon bolometer at the Institute of Physics of Charles University.
The surface morphology of the samples was measured by atomic force microscopy (AFM) in a contact mode, scanning electron microscopy (SEM), and graphene was characterized using micro-Raman mapping with  532~nm  laser  excitation  (power<20~mW)  in the  backscattering  geometry  with the  objective  (Zeiss,  Germany)  of  numerical  aperture  NA=0.9  and  100$\times$magnification. AFM and Raman mapping are performed on WITec alpha3000 RSA confocal Raman microscope.

\section{Intrinsic plasmons}
\begin{figure}[t!]
\includegraphics[width=1\textwidth]{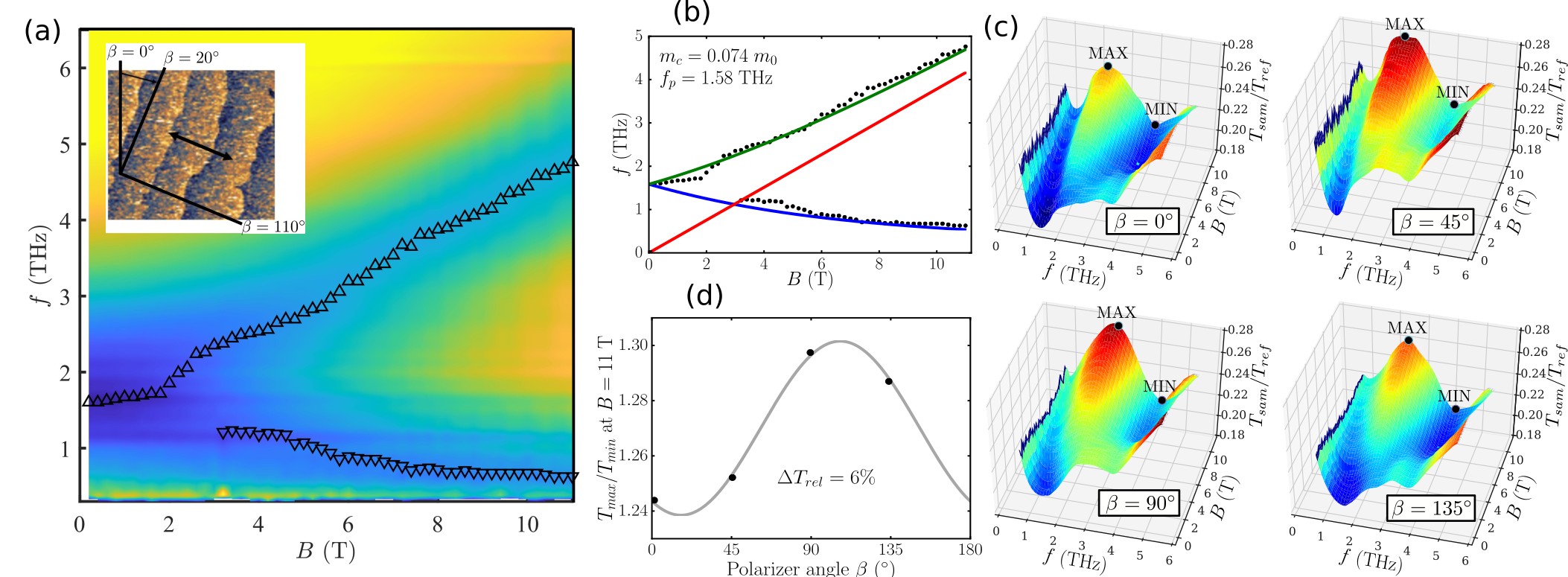}
\caption{Magnetotransmission of Sample~1 in magnetic field up to 11~T (a), inset shows the definition of the angle $\beta$ and the double-arrow depicts polarization of light (electric field) for the maximal visibility of the magneto-plasmon. Fitting of measured (black circles) magnetoplasmon resonance frequencies (b) using eq.~\ref{eq:branches}-green and blue curves for $\omega_+$ and $\omega_-$, respectively.  A straight line is the expected cyclotron resonance with found $m_c=0.074m_e$. Surface plots in (c) show the weak polarization dependence of the transmittance spectra in the magnetic field. Variation of the relative transmittance $\Delta T_{rel}$ with polarizer angle is depicted on (d).}
\label{fig:mplasmon}
\end{figure}

\begin{figure}[t!]
\includegraphics[width=1\textwidth]{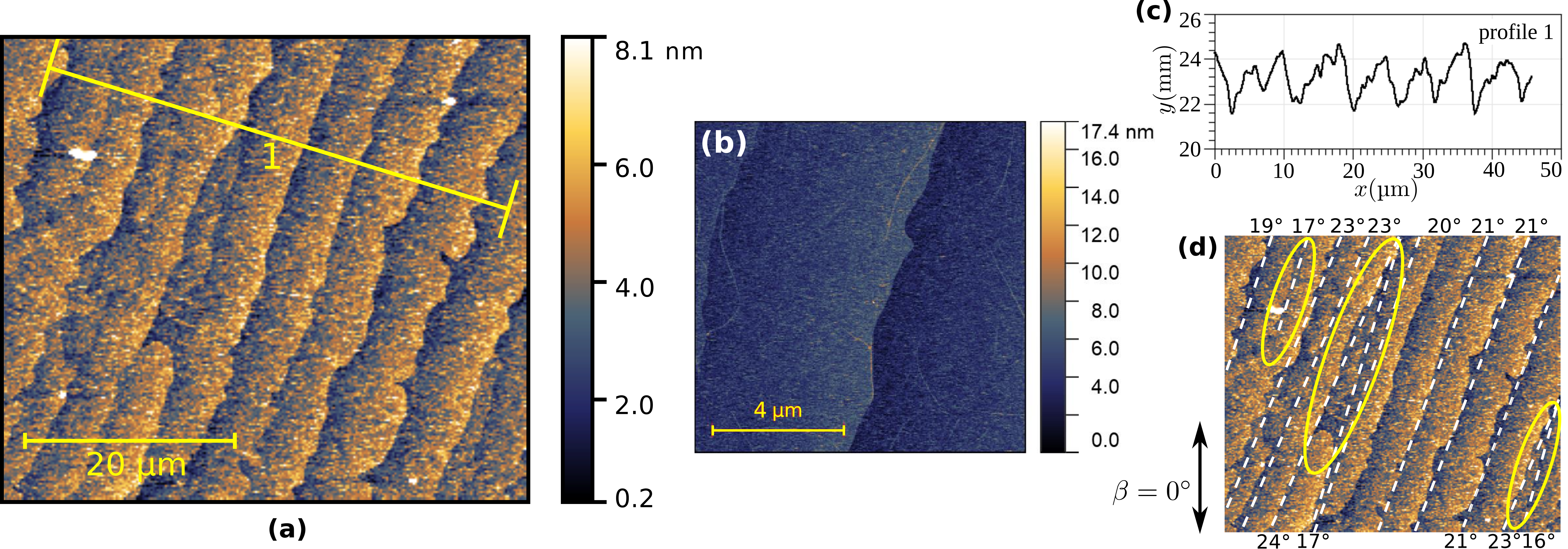}
\caption{Surface topography obtained by AFM showing substrate terraces (a) and detailed map (b). Height profile of the substrate steps from (a) along the yellow line is plotted in (c). Highlighted sub-grain boundaries (yellow ellipses) and terraces (dashed lines) with their angles from the vertical direction (nearest labels) in (d). }
\label{fig:afm}
\end{figure}

We show the magneto-transmittance spectra of QFSLG in Fig.~\ref{fig:mplasmon}~(a). The QFSLG was subjected to a perpendicular magnetic field $B$ up to 11~T. We plot a relative transmittance $T_{M,rel}(\omega, B)$
\begin{equation}
T(\omega, B)_{rel}=\frac{T(\omega,B)_{sam}/T(\omega,0)_{sam}}{T(\omega,B)_{sub}/T(\omega,0)_{sub}},
\label{eq:relT}
\end{equation}
where $T(\omega,B)_{sam}/T(\omega,0)_{sam}$ and $T(\omega,B)_{sub}/T(\omega,0)_{sub}$ correspond to the relative transmittance of QFSLG and bare SiC reference sample related to zero magnetic field. The latter ratio also includes the effects of the magnetic field-dependent sensitivity of the bolometer and other field-induced effects independent of the sample. Hence, the ratio in Eq.~(\ref{eq:relT}) represents a true relative (to 0~T) change in graphene's transmission induced by the magnetic field.
\\
Despite the lack of artificial structuring, we already observe the magneto-transmission with two minima typical for magneto-plasmon resonance, as shown in Fig.~\ref{fig:mplasmon}~(a).
\\
\\
The local minima from Fig.~\ref{fig:mplasmon}~(a) are related to the confined plasmon $B$-field splitting. The spectral positions of upper $(\omega_+)$ and  lower $(\omega_-)$ branches are shown in Fig.~\ref{fig:mplasmon}~(b) and fitted using the equation~\cite{Allen1983}
\begin{equation}
\omega_\pm=\sqrt{\frac{\omega_c^2}{4}+\omega_p^2}\pm\frac{\omega_c}{2}.   
\label{eq:branches}
\end{equation}
The cyclotron frequency $\omega_c={eB}/{m}=eBv_F^2/E_F$ is given by effective mass $m$, elementary charge $e$ and plasmon resonance frequency $\omega_p$ at $B=0~$T. 
The fit of the experimental data according to the Eq.~\ref{eq:branches} yields the plasmon resonance frequency $f_p=1.6$~THz and cyclotron mass $m=0.074m_0$ in which $m_0$ is the electron rest mass. Considering the Fermi velocity in graphene~\cite{Yin2014a} $v_F\approx 10^6~\mathrm{ms^{-1}}$, $\left|E_F\right|=mv_F^2$ yields the Fermi level of $\left|E_F\right|=420$~meV.
\\
Fig.~\ref{fig:mplasmon}~(c) shows the polarization-dependent transmittance of QFSLG in the magnetic field given by $T(\omega,B)_{sam}/T(\omega,B)_{ref}$ for four polarization angles with a step of $45^\circ$ covering the interval of $180^\circ$. At first glance, the transmittance seems to be polarization independent. To highlight this, Fig.~\ref{fig:mplasmon} (d) shows the angular dependence of the ratio of maximal (MAX) and minimal (MIN--upper branch of magnetoplasmon) transmittances $T_{max}$ and $T_{min}$, respectively, at $B=11~\mathrm{T}$ from Fig.~\ref{fig:mplasmon}~(c). Here we observe weak transmittance modulation with an amplitude of $\Delta T_{rel}$ of $\sim 6\%$. The maximum effect of magneto-plasmon absorption estimated by squared trigonometric sinus interpolation was found for $\beta\approx 110^\circ$, which means that the coupling of the radiation to the plasmon is stronger for the electric field component perpendicular to the edges of SiC terraces. A similar magneto-plasmon was observed on a non-structured epitaxial graphene sheet by Crassee et {\it al.}~\cite{Crassee2012}. The results are similar concerning the zero-field resonance position, field dependence, and the coupling strength of the plasmon to the radiation with an electric field parallel and perpendicular to SiC crystallographic steps. Within the measurement uncertainty, we do not observe any shift of the plasmon zero-field frequency with the polarization of the impinging light.
\\
Fig.~\ref{fig:afm}~(a) shows the surface topography covering the area of $50\times 50~\mathrm{\mu m^2}$. Here a series of well-defined parallel substrate terraces can be observed (Fig.~\ref{fig:afm}~(b) in detail), whose widths vary between approximately 5 and $7~\mathrm{\mu m}$ with the average step height of about 2~nm (see profile one on Fig.~\ref{fig:afm}~(c)). We analyzed the shapes of substrate terraces steps by approximating them by dashed lines in Fig.~\ref{fig:afm} from which we can observe some of their features, especially several dislocations marked by yellow ellipses and their varying slopes (dashed guidelines in Fig.~\ref{fig:afm}~(c)). The analysis results in the angle of $\beta=(20\pm3)^\circ$ between the terraces' steps and the vertical direction. This angle coincides with the minimum of the coupling strength of the radiation to the plasmons, Fig.~\ref{fig:mplasmon}~(d). We stress that the absorption of a confined plasmon mode should vanish completely when $\mathbf{E}$ is parallel with the steps of the terraces, while in reality, its amplitude varies by only 6\% with the polarizer angle. Since our QFSLG exhibits a higher degree of surface anisotropy than in Ref.~\cite{Crassee2012}, and we observe (within an experimental error) the same coupling strength variation (6\%) as in Ref.~\cite{Crassee2012}, we conclude that the SiC step edges and crystallographic terrace orientation are not responsible for the intrinsic spatial confinement in QFSLG.

\begin{figure}[h!]
\centering
\includegraphics[width=0.65\textwidth]{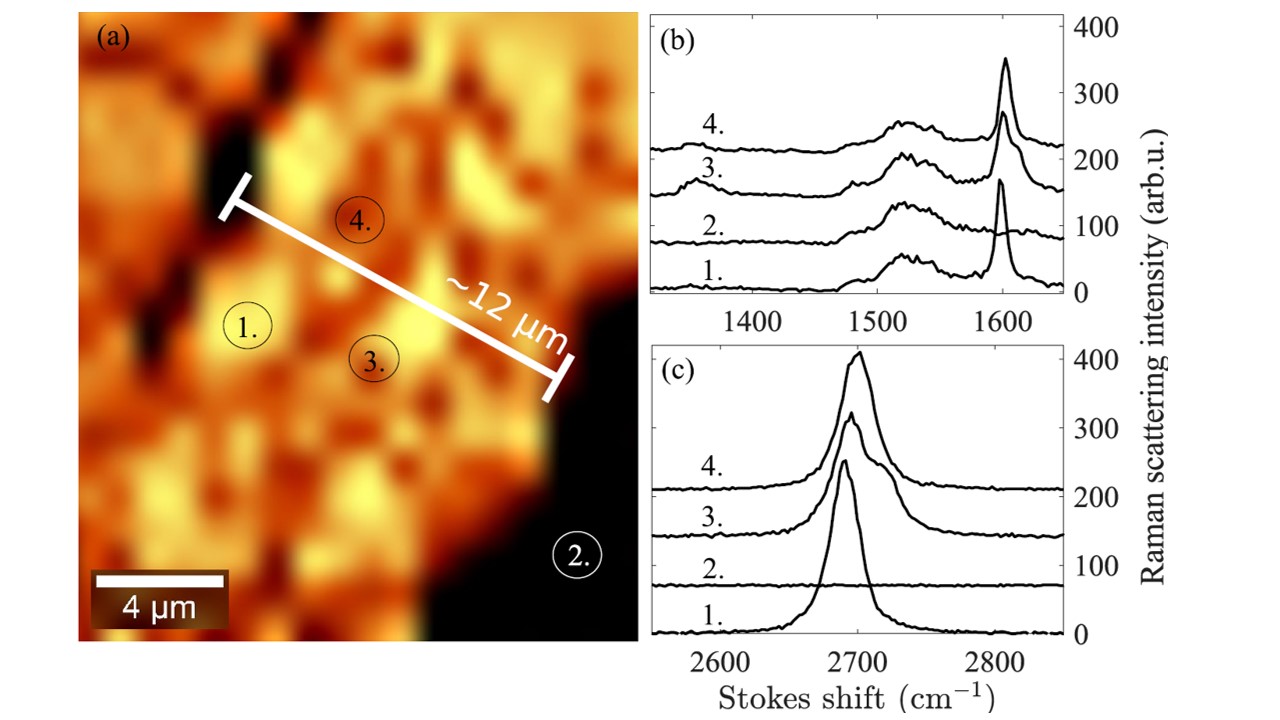}
\caption{Micro Raman map focused on the detail of QFSLG (a). Plots (b) and (c) show representative micro-Raman spectra of four chosen spots in the vicinity of graphene G-peak and 2D-peak, respectively.}
\label{fig:raman}
\end{figure}
To further investigate the structure of QFSLG, we show the micro-Raman mapping in Fig.~\ref{fig:raman}. Most of the surface is formed by QFSLG. Dark spots mark areas without graphene and the boundaries between its flakes. We observe approximately $12~\mathrm{\mu m}$ wide graphene structures, which is in good agreement with the effective dimension of graphene flakes obtained below from analysis of the plasmonic absorption measurements.
Comparing Figs.~\ref{fig:afm} (AFM) and~\ref{fig:raman} (Raman map), we conclude that the dimensions of graphene flakes are primarily independent of the surface morphology. However, the graphene grains are weakly elongated along the SiC terraces. This weak elongation leads to the weak modulation of the polarization dependence ($\Delta T_{rel.}=\left(T_{min}/T_{max}\right)=6\%$ at $B=11~\mathrm{T}$) of the plasmon resonance in the as-grown QFSLG. Thus, the high degree of the SiC terraces' anisotropy and the weak plasmon resonance polarization anisotropy exclude the terraces from being the origin of the translational symmetry breaking. Also, the two observed magneto-plasmon branches correspond to the two-dimensional confinement, whereas the SiC terraces should exhibit one-dimensional confinement, showing only one magneto-plasmon branch. 
The unknown spatial confinement length is the major drawback in determining the origin of the plasmon resonance in the intrinsic as-grown graphene. The Raman mapping in Fig.~\ref{fig:raman} gives only a local estimate. Otherwise, the confinement length remains a fitting parameter. 

\section{Plasmon in GNR at $B=0$~T}
In the following section, we confront the above conclusions with the data collected on artificially made graphene nanoribbons (GNR). The advantage of the lithographically-defined GNRs is the well-defined GNR's width, spacing, and Fermi level. We determine the dimensions by Scanning Electron Microscopy (SEM), AFM, and the Fermi level by measuring the plasmon resonance in the magnetic field (magneto-plasmon). We studied three GNR samples. Two are made of single-layer graphene (SLG), and one sample consists of the hydrogen-intercalated buffer, the quasi-free standing single-layer graphene (QFSLG). 

\begin{figure}[t!]
\centering
\includegraphics[width=0.7\textwidth]{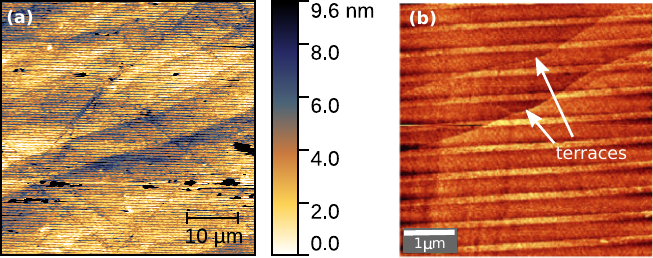}
\caption{(a) Surface topography of GNRs acquired by AFM showing the mutual orientation of lithographic stripes and substrate terraces and (b) higher resolution map.}
\label{fig:afm2}
\end{figure}

We commence with the argon-grown GNRs. The large- and small-area AFM in Fig.~\ref{fig:afm2} shows the ribbons and the SiC terraces. Substrate surface terraces intersect GNRs at an angle of approximately $20^\circ$.

\begin{figure}[ht!]
\includegraphics[width=1\textwidth]{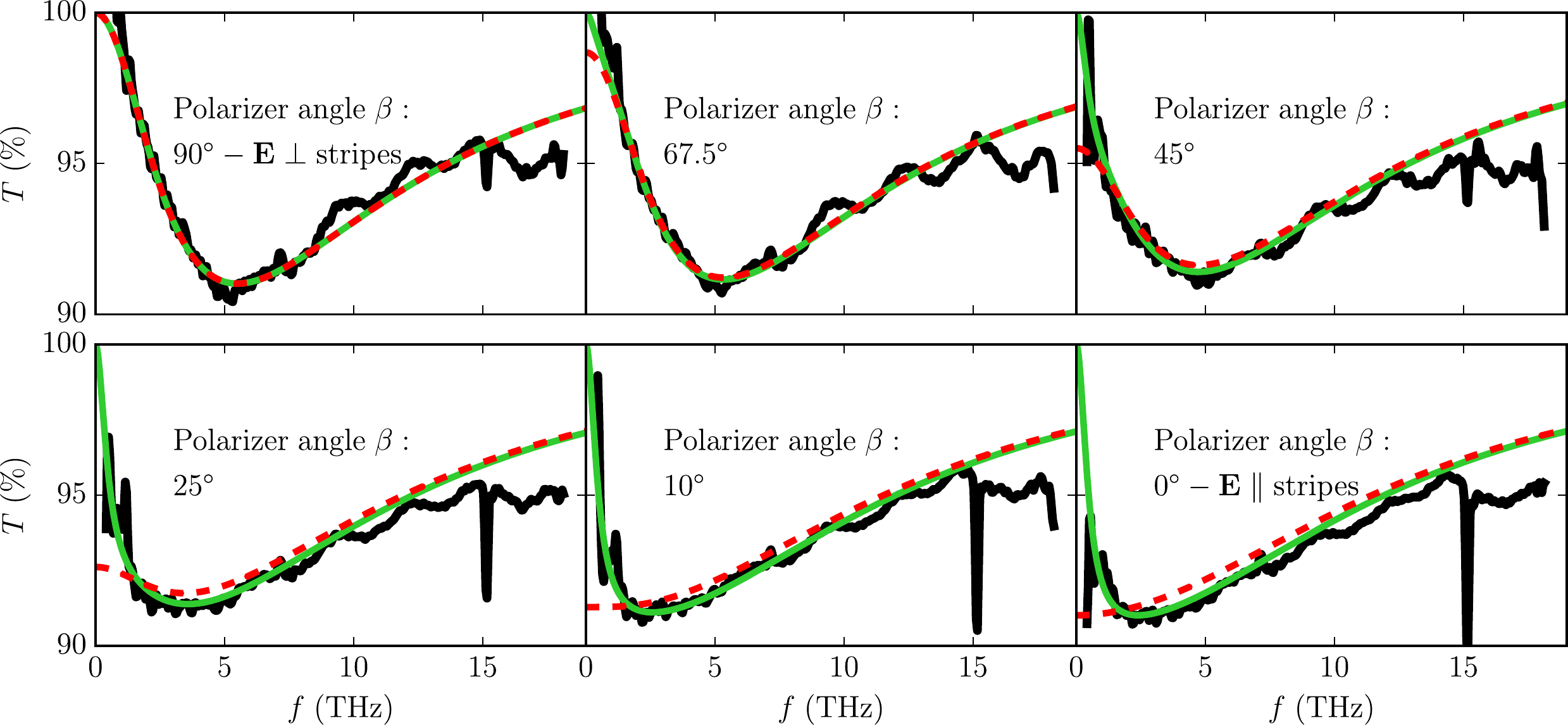}
\caption{FIR transmittance of Sample 2 with lithographically striped SLG graphene (black curves-experimental data) at various polarizer angles ranging between 0 and $90^\circ$. Red dashed curves represent the simulations of the transmittance considering lithographic stripes only (eq.~\ref{eq:pr2}) while the solid green curves show the simulations including the graphene structures along the stripes (eq.~\ref{eq:pr3}).}
\label{fig:figLD}
\end{figure}

We present the FIR optical transmittance at room temperature and $B=0$~T in Fig.~\ref{fig:figLD}. The resonant frequency $f_p\approx5.2(3)$~THz corresponding to the confined plasmon mode in GNR is apparent when the electric field vector $\mathbf{E}$ of the illuminating light beam is perpendicular to ribbons. This high-energy plasmon resonance is visible for polarizer angles ranging between $90^\circ$ and $45^\circ$. For the light polarized along the GNRs, the minimum transmittance appears at a lower frequency of about 2.4 THz. This low-energy resonance is related to the confinement caused by the finite graphene grain size, as discussed in the first part of the paper. 

We used a damped oscillator model~\cite{Yan2013, Strait2013, Nene2014} to fit experimental FIR transmittance spectra $T_G$ of GNRs quantitatively. The data are ratios of GNR and SiC substrate transmittance $T_{sam}$ and $T_{sub}$, respectively:
\begin{equation}
T_G(\omega)=\dfrac{T_{sam}(\omega)}{T_{sub}(\omega)}=\left|1+\dfrac{e^2}{\pi\hbar^2}\dfrac{Z_0FE_F}{1+n_{SiC}}\dfrac{i\omega}{\omega^2-\xi\omega_p^2+i\omega/\tau}\right|^{-2}.
\label{eq:pr1}
\end{equation}

Here $Z_0$ is the impedance of vacuum, $F=W/R$ is the filling factor---the fraction of graphene covered area, GNR width $W$, GNR array period $R$, $n_{SiC}=2.59$ is the substrate refractive index, and $\omega_p=2\pi f_p$ is the plasmon resonance frequency. $\tau$ is the carrier relaxation time. A discrete factor $\xi=1$ if the stripes are perpendicular to $\mathbf{E}$ and $\xi=0$ if they are parallel.

The spectral dependence of $T(\omega)$ on the polarizer angle  $\beta$ was simulated by the following combination of both Lorentz (plasmon) and Drude (free carriers) transitions 
\begin{equation}
T(\omega,\beta)=
\sin^2\beta\left|1+\dfrac{e^2}{\pi\hbar^2}\dfrac{Z_0FE_F}{1+n_{SiC}}\dfrac{i\omega}{\omega^2-\omega_p^2+i\omega/\tau}\right|^{-2}
+\cos^2\beta\left|1+\dfrac{e^2}{\pi\hbar^2}\dfrac{Z_0FE_F}{1+n_{SiC}}\dfrac{i\omega}{\omega^2+i\omega/\tau}\right|^{-2}.
\label{eq:pr2}
\end{equation}
Simulations according to the Eq.~(\ref{eq:pr2}) are plotted as red dashed curves in Fig.~\ref{fig:figLD}. We achieve relatively good agreement with experimental data for $\beta\in(90^\circ,~45^\circ)$ in the spectral range 2.5-15~THz. The simulation yields the carrier relaxation time $\tau=13$~fs. The model becomes more Drude-like for lower $\beta$. However, experimental data for the frequencies below 2~THz still appear to increase towards unity at 0~THz at lower $\beta$. Existing confinement of graphene charges moving parallel to the stripes can only explain this discrepancy. Consequently, to fit the experimental data in Fig.~\ref{fig:figLD}, we also consider a resonance $\omega'_p=2\pi f'_p$ for the motion along the stripes. This model is taken into account in the second term of the following equation:
\begin{equation}
T(\omega,\beta)=
\sin^2\beta\left|1+\dfrac{e^2}{\pi\hbar^2}\dfrac{Z_0FE_F}{1+n_{SiC}}\dfrac{i\omega}{\omega^2-\omega_p^2+i\omega/\tau}\right|^{-2}
+\cos^2\beta\left|1+\dfrac{e^2}{\pi\hbar^2}\dfrac{Z_0FE_F}{1+n_{SiC}}\dfrac{i\omega}{\omega^2-\omega'^2_p+i\omega/\tau}\right|^{-2}.
\label{eq:pr3}
\end{equation}
Polarization-resolved transmittance simulated by Eq.~(\ref{eq:pr3}) is plotted as solid green curves in Fig.~\ref{fig:figLD}. This model fits better experimental data than the simple one described by Eq.~(\ref{eq:pr2}), containing only the lithographically made GNRs. This good agreement supports the hypothesis of graphene structures with two different perpendicular effective dimensions. The similar angular dependence of plasmonic absorption on separate elliptical graphene nanodiscs was described in the work of Xia et al.~\cite{Xia2019}. The second resonance frequency is $f'_p=\omega'_p/2\pi=2.4$~THz with the lifetime $\tau=10$~fs. The Fermi level $E_F = 380-420$~meV is the only parameter determining the amplitude of plasmon absorption.
\\
However, as the ribbon width is determined lithographically ($W=370$~nm), the Fermi level is also unambiguously determined by the spectral position of plasmon resonance. The relation between the small magnitude wave vectors $q=\pi/W$~\cite{Yan2013} and resonance frequency $f_p$~\cite{Strait2013,  Vasic2013, Goncalves2016} follows from the plasmon dispersion
\begin{equation}
f_p=\dfrac{e}{h}\sqrt{\dfrac{E_Fq}{2\pi\epsilon}}.
\label{eq:relpr}
\end{equation}
The permittivity $\epsilon=\epsilon_0\overline{\epsilon}=5.3\epsilon_0$ is taken as an average between the SiC substrate and vacuum/air surrounding graphene. Surprisingly, the Fermi level $E_F = 110$~meV determined from the position of plasmon resonance is in direct contradiction with the Fermi level determined from the absorption intensity. The Fermi level $E_F = 110$~meV is unusually smaller than the one measured on our samples by transport Hall effect measurements~\cite{KuncPRA2017}, and it is also smaller than typical results reported in the literature~\cite{Mammadov2DMaterials2017}. To unambiguously determine the Fermi level in our samples, we measured the plasmon mode coupled to the magnetic field (magneto-plasmon), where the Fermi level determines the high-field slope of the magneto-plasmon $B$-dependent resonance.

\section{Magneto-plasmon in GNR}
\begin{figure}[t!]
\centering
\includegraphics[width=0.8\textwidth]{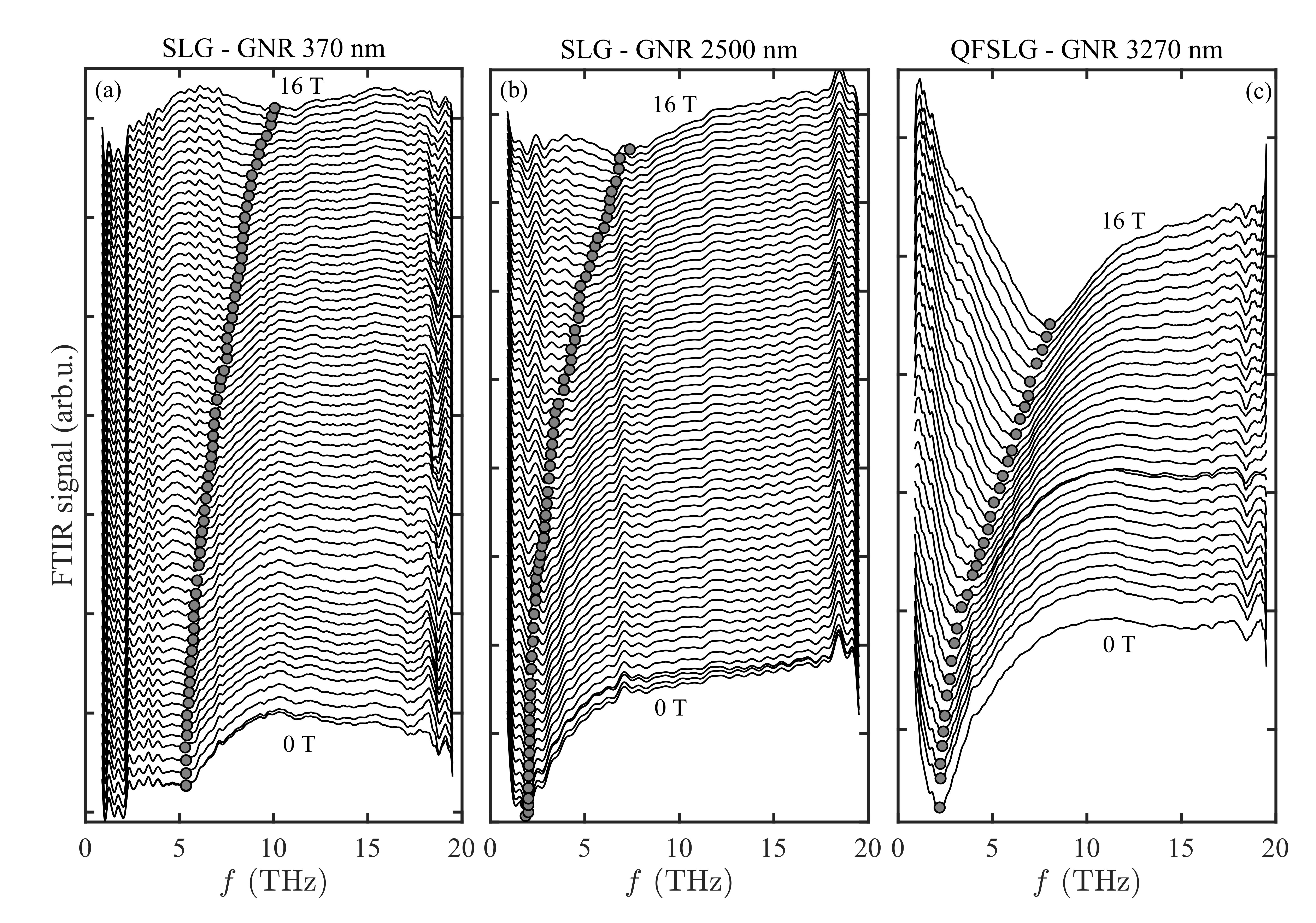}
\caption{Magneto-plasmon spectra of (a) 370~nm, (b) 2500~nm and (c) 3270~nm wide graphene nanoribbons. The ribbons are separated by (a) 130~nm, (b) 500~nm, and (c) 540~nm.}
\label{fig:waterfall}
\end{figure}

\begin{figure}[t!]
\centering
\includegraphics[width=0.35\textwidth]{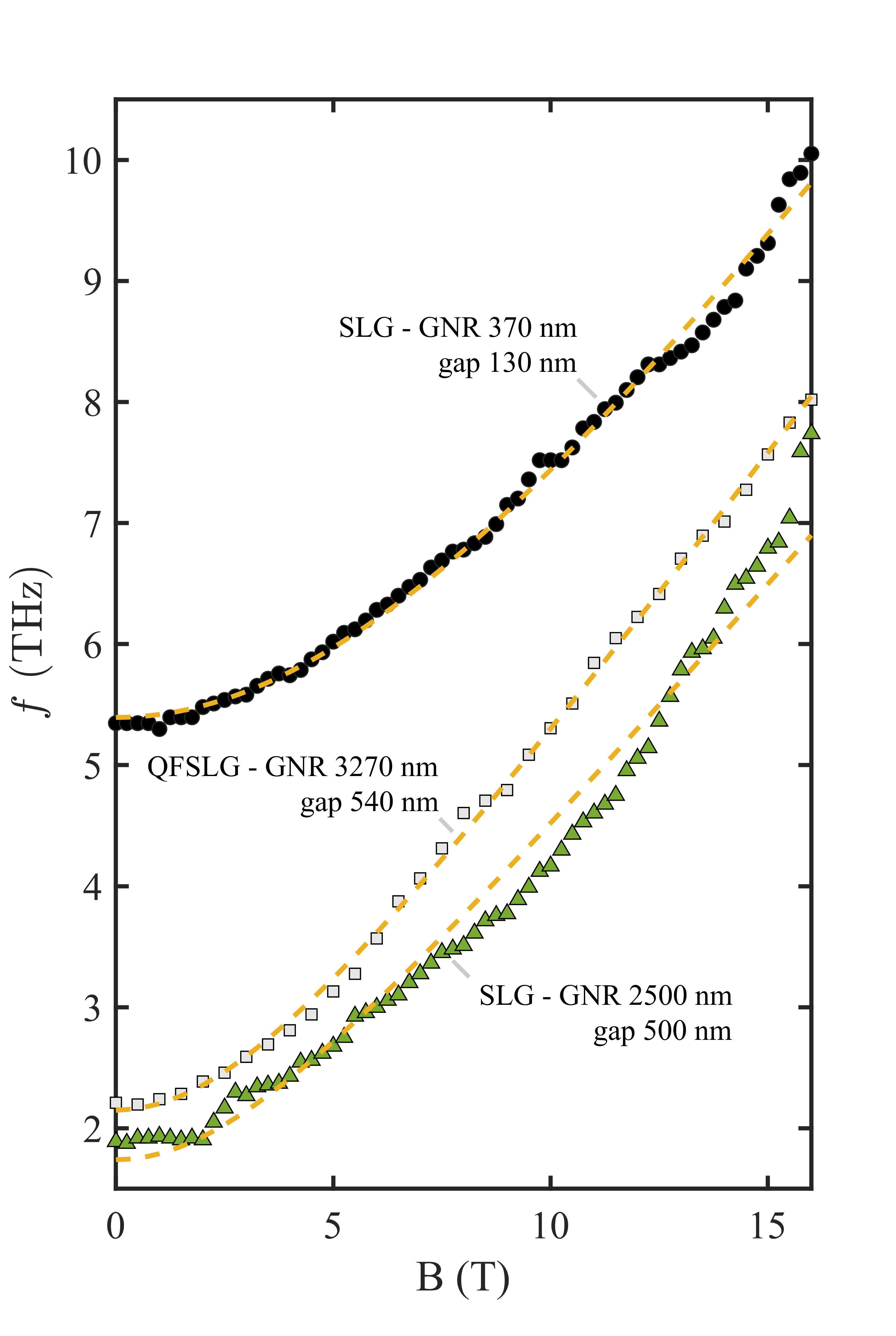}
\caption{Magneto-plasmon peak position of (black points) 370~nm,  (green triangles) 2500~nm, and (grey squares) 3270~nm wide graphene nanoribbons. The yellow dashed lines are Eq.~(\ref{eq:MPPpositionGNR}) fits.}
\label{fig:GNRplasmonPosition}
\end{figure}

We present magneto-transmission data in Fig.~\ref{fig:waterfall} for three GNR samples. The three samples (Sample 2, 3 and 4) have GNR width $W=370$~nm, 2500~nm, and 3270~nm, and the spacing between ribbons is $\omega_{gap}=130(10)$~nm, 500(10)~nm, and 540(5)~nm, respectively. The spectral position of the confined magneto-plasmon mode is shown in Fig.~\ref{fig:GNRplasmonPosition} for magnetic fields up to 16~T. 
The position $\omega_\mathrm{mp}$ of the magneto-plasmon mode in GNR is given by~\cite{Goncalves2016,PoumirolPRL2013,DemelPRB1988}
\begin{equation}
    \omega_\mathrm{mp}=\sqrt{\omega_\mathrm{0,mp}^2+\left(\frac{eBv_F^2}{E_F} \right)^2}.
    \label{eq:MPPpositionGNR}
\end{equation}
The first term determines the zero-field plasmon mode position $\omega_\mathrm{0,mp}=2\pi f_\mathrm{0,mp}$. The second term describes the magnetic-field dependence of the magneto-plasmon mode and allows us to determine the Fermi level of the sample. We can notice that the observed higher plasmon energy in wider ribbons (3270~nm wide), compared to the plasmon energy in 2500~nm wide GNR, is in contradiction with the plasmon dispersion, Eq.~(\ref{eq:relpr}). The critical reason for this difference is, as we discuss later, hydrogen intercalation. The 2500~nm wide GNRs are made of argon-grown graphene without hydrogen intercalation, and we will show later that the presence of the buffer layer leads to a significant redshift of the plasmon mode resonance. We summarize the samples' properties and the fitted $f_\mathrm{0,mp}$ and $E_F$ in Tab.~\ref{tab:samres}.
\begin{table}[h!]
\centering
\caption{Properties of the GNR samples. The GNRs' widths and gap sizes were determined by SEM and AFM. The Fermi level and the zero-field resonance are determined from the magneto-plasmon resonance fitted by Eq.~(\ref{eq:MPPpositionGNR}).}
\begin{tabular}{l c c c} 
 \hline
 \hline
 Sample No.   & Sample 2 & Sample 3 & Sample 4 \\
 Sample structure   & SLG-GNR 370 nm & SLG-GNR 2500 nm  & QFSLG-GNR 3270 nm \\ [0.5ex] 
 \hline
 \multicolumn{4}{c}{SEM+AFM}\\
\hline
Ribbon width $W$    & 370(30)~nm& 2500(100)~nm& 3270(70)~nm\\
Gap between ribbons $\omega_\mathrm{gap}$    & 130(10)~nm& 500(10)~nm& 540(5)~nm \\    
\hline
 \multicolumn{4}{c}{Magneto-plasmon}\\
\hline
$E_F$    &311(2)~meV& 381(4)~meV&  328(2)~meV\\
$f_\mathrm{0,mp}$    &5.39(2)~THz& 1.74(6)~THz& 2.15(2)~THz \\
\hline
\end{tabular}
\label{tab:samres}
\end{table}
In the next section, we develop several models predicting the zero-field plasmon spectral position $f_0$ as a function of independently experimentally determined parameters (GNR widths $W$, gaps $\omega_\mathrm{gap}$ and Fermi levels $E_F$). Confrontation of these models with the experimental data will allow us to develop our interpretation of the observed phenomena.

\section{Models}
\begin{figure}[t!]
\centering
\includegraphics[width=1\textwidth]{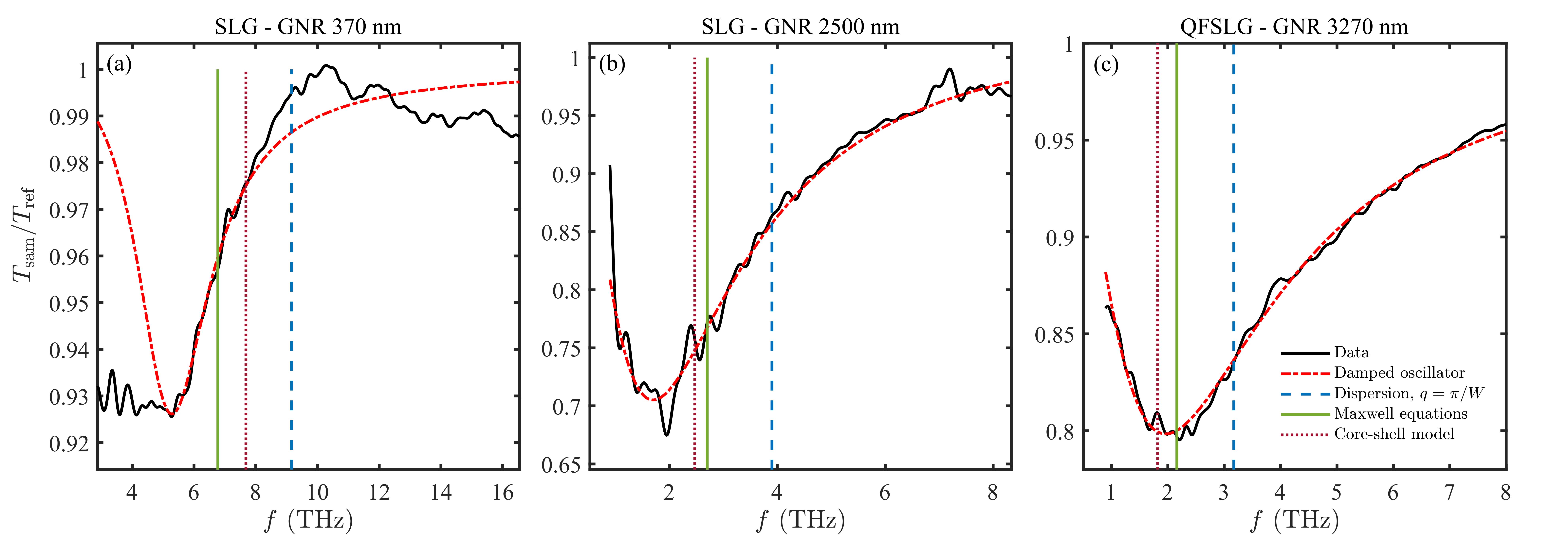}
\caption{Polarization unresolved plasmon resonances are compared to the fits by a damped oscillator model, shown by the dash-dot red curves. The plasmon resonance positions are shown for the models based on (dashed blue line) the plasmon dispersion, (dotted red line) core-shell model, and (solid green line) exact solution of Maxwell equations. }
\label{fig:DataAllModelsCompared}
\end{figure}
We compare three models to evaluate the plasmon mode energy from the known GNR dimensions and the Fermi energy. These models are based on the plasmon dispersion (Disp.), core-shell model (CS.), and exact numerical solution of Maxwell equations (ME.). We then introduce the calculated plasmon mode frequencies into a damped oscillator model (DOM), Eq.~\ref{eq:pr1}, to compare the resulting spectrum with the measured data. For comparison, we also fit DOM to the data considering the plasmon frequency $f_\mathrm{0,dom}$ as a fitting parameter. The relevant parameters of these calculations are summarized in Tab.~\ref{tab:models}.

\subsection{Plasmon dispersion}
The plasmon dispersion predicts the zero-field plasmon mode frequency given by Eq.~(\ref{eq:relpr}). The Fermi energy and the wave vector $q$ are the input quantities. The basic model gives $q=\frac{\pi}{W}$, for the GNR width $W$~\cite{MikhailovPRB2005,Yan2013,Goncalves2016}. There is no dependence on the gap between two neighboring ribbons $\omega_\mathrm{gap}$~\cite{Yan2013}. We take the experimental values of the GNR widths $W$ and Fermi energies determined from the magneto-plasmon, and we summarize the resulting plasmon energies in Tab.~\ref{tab:models}. This model overestimates the plasmon mode frequency for all three GNR samples. We depict the overestimated plasmon frequencies by dashed blue vertical lines in Fig.~\ref{fig:DataAllModelsCompared}. 

\subsection{Core-shell model}
The two-dimensional analog of the core-shell (CS) model~\cite{Landau,Rychetsky2020,PaingadAFM2021} effectively includes the plasmon-plasmon interaction. The model provides correction of the wave vector $q=\pi/W$ to be also given by the gap between graphene ribbons $\omega_\mathrm{gap}$ 
\begin{equation}
    q=\frac{2\pi\omega_\mathrm{gap}}{W^2}.
\label{eq:core}    
\end{equation}
The core-shell model predicts redshifted plasmon mode resonances for all three samples. However, the predicted redshift for samples 2 and 3 is not sufficient. On the other hand, the redshift for the hydrogen-intercalated GNR, sample 4, is slightly overestimated. Generally, the agreement with experimental data is better than in the case of the bare plasmon dispersion, but it is still not quantitative. We depict the plasmon mode frequencies predicted by the core-shell model by dotted red vertical lines in Fig.~\ref{fig:DataAllModelsCompared}. 
\begin{figure}[t!]
\centering
\includegraphics[width=1\textwidth]{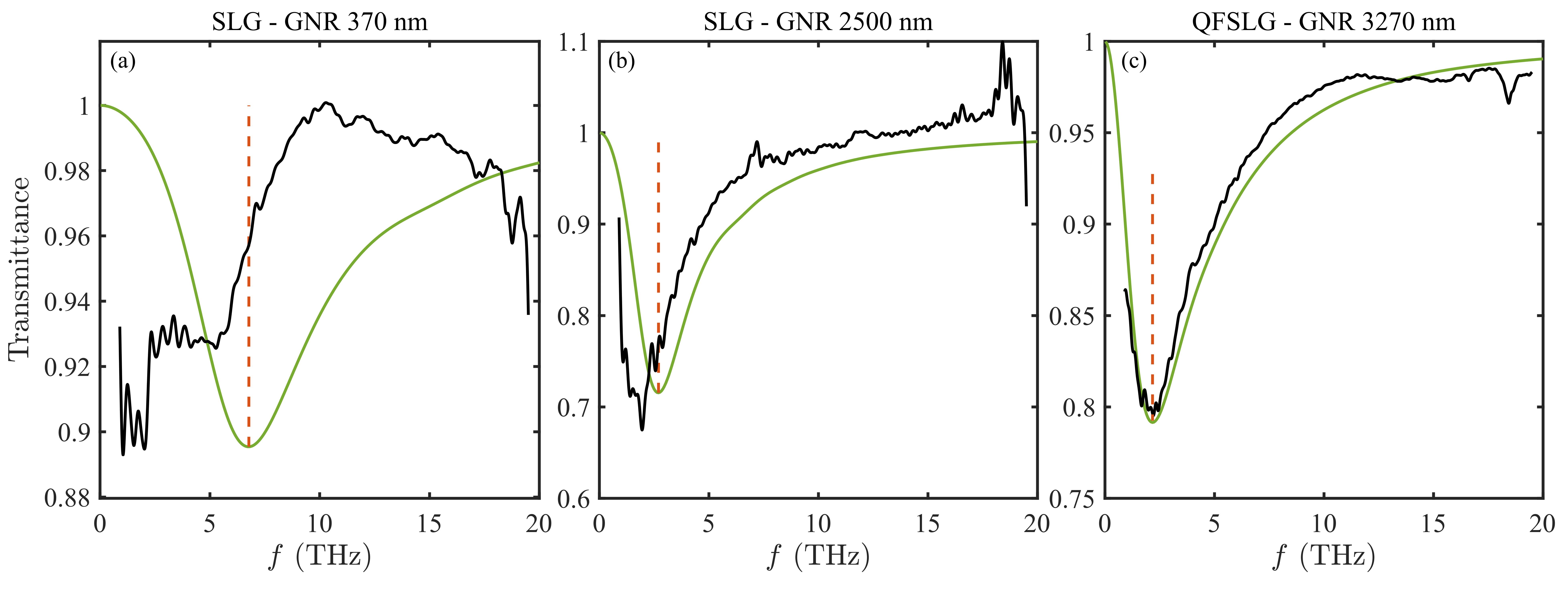}
\caption{The experimental data (black curves) are compared with the exact solution of Maxwell equations (green curves) for the three GNR samples.}
\label{fig:MaxwellExact}
\end{figure}
\subsection{Maxwell equations numerically}
We attempted to understand the quantitative disagreements of the above models by an exact solution of Maxwell equations. The periodicity of the GNR array allows expanding the electromagnetic field in Bloch-type surface waves~\cite{Goncalves2016}, p.168-174. We solve the boundary value problem as a scattering problem, where the impinging radiation is a source of the scattered field. We assume Drude-like graphene conductivity. The input parameters are the GNR width, the gap between ribbons, Fermi energy, and the relaxation time $\tau$. The relaxation time is the only variable parameter here.\\
We compare the best-modeled spectra with the experimental data in Fig.~\ref{fig:MaxwellExact}. The simulated spectra of non-intercalated GNRs in samples 2 and 3 significantly disagree with the measured spectra. However, the hydrogen intercalated GNR, sample 4, shows perfect agreement with the data. The experimentally measured plasmon mode frequency is 2.2(1)~THz, and the numerically determined value is the same, including error. The error in the numerically simulated spectrum is evaluated by running the numerical simulation for the upper/lower bound of the Fermi energy and the lower/upper bound of the GNR width. Following Fig.~\ref{fig:DataAllModelsCompared} and Tab.~\ref{tab:models}, the ME model provides the closest match to the DOM fit of the data for all the samples. While for QFSLG-GNR (sample 4) ME model shows a perfect agreement with the data, the simulated spectra for non-intercalated SLG-GNRs (samples 2 and 3) still disagree with the experiment (see Fig.~\ref{fig:MaxwellExact}). The well-predicted plasmon mode frequency in the QFSLG-GNR leads us to conclude that the plasmon-plasmon interaction is behind the redshift in the hydrogen-intercalated sample. 
\begin{table}[t!]
\centering
\caption{Calculated plasmon frequency at zero magnetic field using the three models: Disp., Eq.~(\ref{eq:relpr}), taking $q=\pi/W$. The CS model $f_\mathrm{0,cs}=f_\mathrm{0,disp}\sqrt{2\omega_\mathrm{gap}/W}$, and ME, where the relaxation time $\tau$ was fitted to the broadening of the zero-field plasmon mode. The values are compared with an ad hoc fit by DOM, Eq.~(\ref{eq:pr1}).}
\begin{tabular}{l l c c c} 
 \hline
 \hline
 Model&   & SLG - GNR 370 nm & SLG - GNR 2500 nm & QFSLG - GNR 3270 nm \\ 
   & & Sample 2 & Sample 3  & Sample 4 \\
 \hline
Disp. & $f_\mathrm{0,disp}$ &9.2(2)~THz& 3.9(1)~THz& 3.17(4)~THz \\
CS   & $f_\mathrm{0,cs}$   &7.7(5)~THz& 2.5(4)~THz& 1.8(2)~THz \\
ME &$f_\mathrm{0,me}$   &6.8(6)~THz  &2.7(2)~THz  & 2.2(1)~THz \\
   & $\tau$ & 25(5)~fs & 60(5)~fs  & 43(10)~fs \\
\hline
DOM   &$f_\mathrm{0,dom}$& 5.3(1)~THz&1.70(2)~THz& 2.02(2)~THz \\
&$\tau$ &25(15)~fs& 58(5)~fs & 48(5)~fs\\
    &$E_F$&55.3(6)~meV & 301(20)~meV& 180(40)~meV \\
\hline
Plasmon resonance \\($B=0$~T FTIR data) & $f_\mathrm{0,exp}$   &5.2(3)~THz & 1.9(1)~THz& 2.2(1)~THz \\
 \hline
\end{tabular}
\label{tab:models}
\end{table}

\section{Discussion}
\begin{figure}[t!]
\centering
\includegraphics[width=0.5\textwidth]{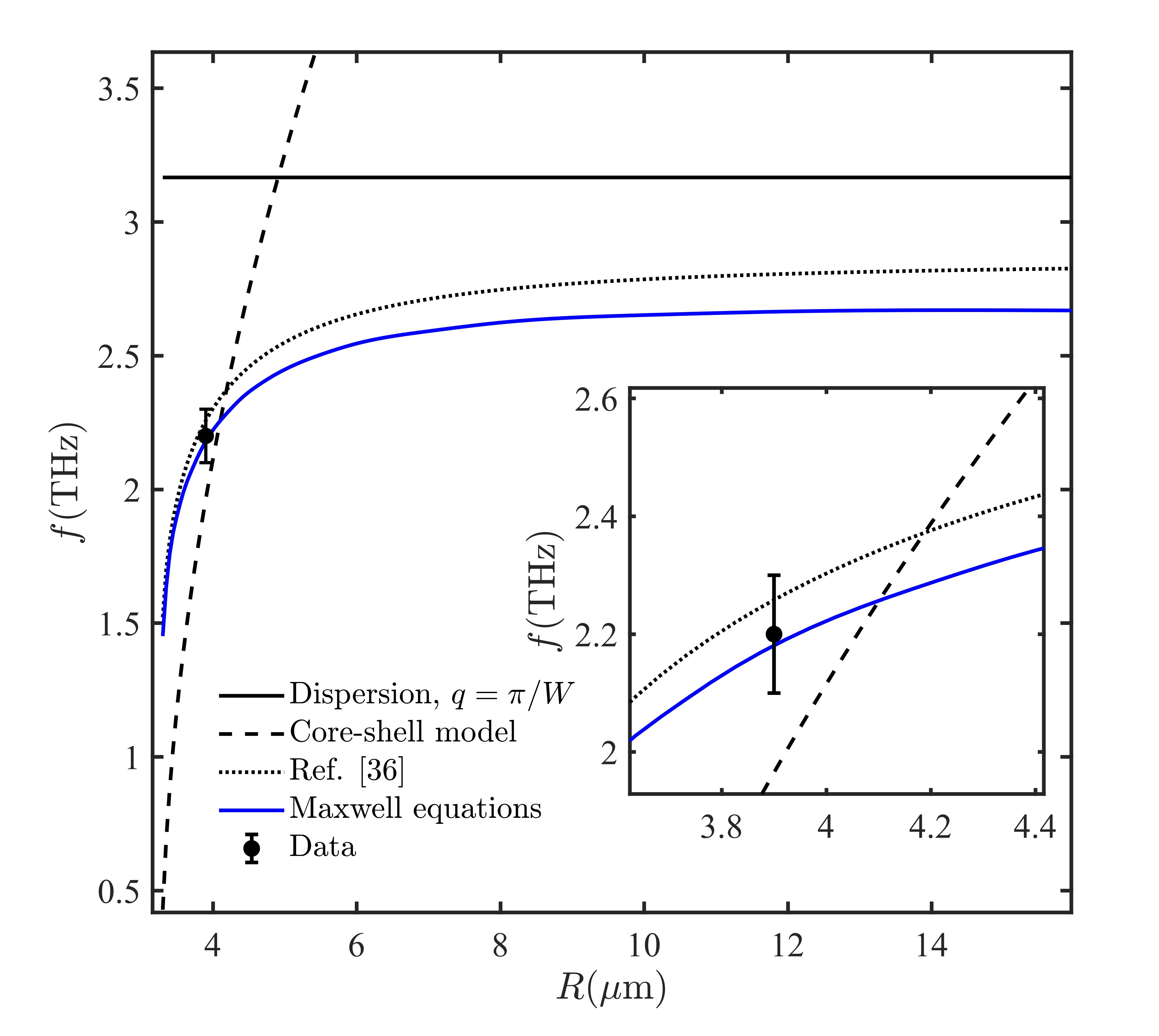}
\caption{The effect of the plasmon-plasmon interaction. The plasmon frequency is plotted versus the graphene array periodicity $R$ for $W=3270$~nm wide GNR array. The plasmon resonance in Sample 4 (QFSLG-GNR 3270~nm), shown by the point with an error bar, is compared with the plasmon resonance predicted by the plasmon dispersion (solid black curve), core-shell model (dashed black curve), equivalent circuit model (Jadidi, Ref.~\cite{JadidiNanoLett2016}, dotted black curve), and the exact solution of Maxwell equations (solid blue curve). }
\label{fig:PlasmonPlasmonInteraction}
\end{figure}
The models used to describe the experimentally observed low-energy plasmon mode resonances gave us essential hints in the data interpretation. Despite providing correct scaling rules ($f\propto E_F^{1/2}$, $f\propto q^{-1/2}$), the basic model only qualitatively describes the plasmon mode frequency. The measured resonance frequency is, in all samples, almost twice smaller. The factor of two in frequency translates into a factor of $\approx 4$ in Fermi energy or the confinement lengths. 
The core-shell model slightly improved the qualitative agreement; however, it does not provide a full quantitative prediction. The exact solution of Maxwell's equations showed the importance of the plasmon-plasmon interaction.\\
We study the plasmon-plasmon interaction in detail in Fig.~\ref{fig:PlasmonPlasmonInteraction}. The solid blue curve shows the numerical result, the plasmon mode transmission minimum, as a function of the nanoribbon array periodicity $R$ for a fixed GNR width $W=3270$~nm. The gap between ribbons $\omega_\mathrm{gap}=R-W$ increases with increasing periodicity $R$. Thus the interaction between plasmons is reduced as $R\rightarrow\infty$. The Fig.~\ref{fig:PlasmonPlasmonInteraction} allows quantifying the spectral redshift caused by the plasmon-plasmon interaction $\Delta f_\mathrm{pp}=f_\mathrm{0,me}^{R\rightarrow\infty}-f_\mathrm{0,me}$. Here $f_\mathrm{0,me}$ is the result of Maxwell equations assuming the true geometry of GNR array, and $f_\mathrm{0,me}^{R\rightarrow\infty}$ is the large-periodicity limit $R\rightarrow\infty$. We also depict the results of the basic plasmon dispersion model and the core-shell model in Fig.~\ref{fig:PlasmonPlasmonInteraction}. We also depict the equivalent circuit model (ECM)~\cite{JadidiNanoLett2016}. The ECM provides a fairly good agreement with the exact solution in the whole range of GNR array periodicities $R$. We note that its prediction is, within an error, in agreement with our experimental data. 
To further elucidate the disagreement between the experimental data and the exact solution for non-intercalated samples, we quantified the plasmon-plasmon interaction also for non-intercalated samples (samples 2 and 3). The results of our numerical calculations are in Tab.~\ref{tab:bufferdiscussion}. 
\begin{table}[h!]
\centering
\caption{Analyses of the plasmon-plasmon interaction redshift, buffer redshift, and the summary of the buffer-related density of states $\gamma$.}
\begin{tabular}{c c c c} 
 \hline
 \hline
    & SLG - GNR 370 nm & SLG - GNR 2500 nm & QFSLG - GNR 3270 nm \\ [0.5ex] 
 \hline
No interaction limit $f_\mathrm{0,me}^{R\rightarrow\infty}$& 7.7(3) THz & 3.2(1) THz & 2.66(3) THz \\ 
plasmon-plasmon\\ interaction $f_\mathrm{0,me}^{R\rightarrow\infty}-f_\mathrm{0,me}$ & 0.9 THz & 0.5 THz &  0.5 THz\\
Buffer redshift $f_\mathrm{0,me}-f_\mathrm{0,exp}$&1.6 THz & 0.8 THz &  0 THz\\
Buffer DoS $\gamma$ & 5.6$\times 10^{12}~\mathrm{cm^{-2}eV^{-1}}$ & 8.2$\times 10^{12}~\mathrm{cm^{-2}eV^{-1}}$ & 0~$\mathrm{cm^{-2}eV^{-1}}$ \\
 \hline
\end{tabular}
\label{tab:bufferdiscussion}
\end{table}
The interaction causes a redshift of 0.9 and 0.5~THz, which cannot explain the total experimentally observed redshift 2.5 and 1.3~THz, respectively (see also Fig.~\ref{fig:DataAllModelsCompared}~(a,b)). The major difference between the intercalated and non-intercalated samples is the presence of the buffer layer. The buffer layer contains a finite density of states which can get charged and, consequently, modify the plasmon resonance~\cite{SasakiPRB2014}. We follow the model of Sasaki and Kumada~\cite{SasakiPRB2014}, replacing the charge $e$ by an effective charge $e_*$, estimated by Luryi~\cite{LuryiAPL1988} $e_*\simeq \frac{C_d}{C_i+C_d}e$. The quantum capacitance $C_i=e^2\gamma$ and the geometrical capacitance $C_d=\epsilon_0/d$ are given by the density of states $\gamma$ and by the buffer-graphene distance $d=0.3$~nm. Assuming that the plasmon frequency is proportional to the effective charge $e^*$, Eq.~(\ref{eq:relpr}), we get for the red-shift due to the buffer states
\begin{equation}
    \frac{f_\mathrm{0,exp}}{f_\mathrm{0,me}}=\frac{C_\mathrm{d}}{C_\mathrm{i}+ C_\mathrm{d}}
\end{equation}
and we can evaluate the buffer density of states $\gamma$ for the two SLG samples,
\begin{equation}
    \gamma=\frac{\epsilon_0}{de^2}\left(\frac{f_\mathrm{0,me}}{f_\mathrm{0,exp}}-1\right).
\end{equation}

Tab.~\ref{tab:bufferdiscussion} summarizes the resulting buffer density of states 5.6 and 8.2$\times 10^{12}~\mathrm{cm^{-2}eV^{-1}}$ for the two non-intercalated samples 2 and 3, respectively. These densities agree with the experimentally reported values 5-10$\times 10^{12}~\mathrm{cm^{-2}eV^{-1}}$~\cite{KopylovAPL2010,JanssenPRB2011,TakasePRB2012}.

Let us return to the intrinsic plasmons in unpatterned graphene. We concluded that the translational symmetry breaking in as-grown hydrogen-intercalated samples is not caused by the SiC terraces. Instead, it is caused by the finite-size graphene grains. The remaining question is whether we can determine the confinement length for the intrinsic plasmons. Unfortunately, based on the analyses presented above, we cannot distinguish between the large non-interacting and small interacting graphene domains. Both cases will lead to the same plasmon mode resonant frequency. 

Our findings also have consequences on the tunability of graphene plasmons. Since graphene grain size determines the lower bound of the plasmonic resonance, it also determines the upper practical limit of graphene nanoribbon width. The improved tunability range towards far-infrared and THz applications requires improvement in the large-scale homogeneous graphene growth. The plasmon-plasmon interaction also provides another route for spectral tunability of the localized plasmon modes. Besides chemical and electrostatic doping and GNR width, this is the gap between GNRs. The large gap could be used to reach plasmon resonance in the mid-infrared range. The drawback is a small coverage of the sample's surface by GNRs, causing smaller overall absorption. 

\section{Conclusions}
We proposed the origin of the observed THz plasmon and magnetoplasmon in non-patterned  epitaxial graphene by the natural formation of graphene micro-flakes accompanying the graphene growth. These structures give rise to a significant contribution to the plasmonic absorption of patterned graphene as well. The proposed structures are almost independent of the morphology of SiC substrate terraces, and they originate in the graphene growth mechanism. The effective size of graphene flakes depends on epitaxial graphene growth conditions (argon-grown or hydrogen-intercalated graphene). Their small anisotropy is only up to 6\% given by SiC step bunching. This natural formation of micro--flakes should be considered during the design and production of plasmonic components on epitaxial graphene, as they determine the lowest achievable energy for THz applications. We confirmed our conclusions with a detailed study of lithographically defined arrays of graphene nanoribbons. We unified the conflicting conclusions from a spectral lineshape analysis, magneto-absorption, and GNRs' dimensions by exact numerical modeling of the plasmon resonance. We found a significant contribution of the plasmon-plasmon interaction and of the buffer layer charging: both leading to a significant redshift of the plasmon. The quantum capacitance model provided the density of interface states of the buffer layer in agreement with the reported values.

\section*{Conflict of Interest}
There are no conflicts to declare.

\section*{Acknowledgement}
This work was supported by the Czech Science Foundation (GA CR), Project No. 19-12052S. CzechNanoLab project LM2018110 funded by MEYS CR is gratefully acknowledged for the financial support of the sample fabrication at CEITEC Nano Research Infrastructure. The work of PK was supported by the Operational Programme “Research, Development and Education” financed by European Structural and Investment Funds and the Czech Ministry of Education, Youth and Sports (Project No. SOLID21-CZ.02.1.01/0.0/0.0/16\_019/0000760). We also acknowledge the support of the LNCMI-CNRS in Grenoble, a member of the European Magnetic Field Laboratory (EMFL).


\begin{footnotesize}
\bibliographystyle{unsrt}
\bibliography{export.bib}
\end{footnotesize}

\end{document}